\documentclass[12pt,preprint]{aastex}
\newcommand{\chandra}{{\it Chandra}}
\newcommand{\xmm}{{\it XMM-Newton}}
\newcommand{\rosat}{{\it ROSAT}}
\newcommand{\asca}{{\it ASCA}}
\newcommand{\hii}{H~{\small II}}
\newcommand{\hi}{H~{\small I}}
\newcommand{\ha}{H$\alpha$}
\newcommand{\sii}{[\ion{S}{2}]}
\newcommand{\oiii}{[\ion{O}{3}]}

\newcommand{\half}{1/2}

\begin{document}

\title{Supernova Remnants in the Magellanic Clouds. VI. The DEM\,L\,316 Supernova Remnants}
\author{R.~M. Williams \& Y.-H. Chu}
\affil{University of Illinois at Urbana-Champaign, 1002 W. Green St., 
Urbana, IL 61801 USA}
\email{rosanina@astro.uiuc.edu, chu@astro.uiuc.edu}


\begin{abstract}
The DEM\,L\,316 system contains two shells, both with the 
characteristic signatures of supernova remnants (SNRs).  We 
analyze \chandra\ and \xmm\ data for DEM\,L\,316, investigating 
its spatial and spectral X-ray features. Our \chandra\ observations 
resolve the structure of the northeastern SNR (Shell A) as a bright 
inner ring and a set of ``arcs" surrounded by fainter diffuse 
emission. The spectrum is well fit by a thermal plasma model with 
temperature $\sim1.4$ keV; we do not find significant spectral 
differences for different regions of this SNR. The southwestern
SNR (Shell B) exhibits an irregular X-ray outline, with a brighter 
interior ring of emission including a bright knot of emission.  
Overall the emission of the SNR is well described by a thermal 
plasma of temperature $\sim0.6$ keV.  The Bright Knot, however, 
is spectrally distinct from the rest of the SNR, requiring the 
addition of a high-energy spectral component consistent with a 
power-law spectrum of photon index 1.6--1.8.

We confirm the findings of \citet{N+01} that the spectra of these 
shells are notably different, with Shell A requiring a high iron 
abundance for a good spectral fit, implying a Type Ia origin. We 
further explicitly compare abundance ratios to model predictions 
for Type Ia and Type II supernovae. The low ratios for Shell A 
(O/Fe of 1.5 and Ne/Fe of 0.2) and the high ratios for Shell B 
(O/Fe of 30--130 and Ne/Fe of 8--16) are consistent with Type Ia 
and Type II origins, respectively.  The difference between the SNR
progenitor types casts some doubt on the suggestion that these 
SNRs are interacting with one another.

\end{abstract}

\section{Introduction}

The peanut-shaped nebula DEM\,L\,316 was first noted by \citet{MC73} 
to have a high \sii/\ha\ ratio, a typical signature of supernova 
remnants (SNRs).  The authors designated the two lobes of this 
system shells A (the northeastern shell) and B (to the southwest);
following this and subsequent work, we shall use the same 
designations.  \citet{M+83} confirmed that the DEM\,L\,316 system 
had the optical, radio, and X-ray characteristics typical of SNRs.
A multi-wavelength study by \citet[][]{W+97} concluded that the most 
probable scenario for DEM\,L\,316's unique morphology was that it 
consisted of a pair of colliding SNRs. Evidence cited in support of 
this scenario included a change in radio polarization between the two 
shells,  the kinematics of the two shells, multiwavelength morphological
features and similar extinctions. 

Alternate scenarios for the physical association of the two shells include 
(a) that the shells are the result of a single explosion into a bi-lobed 
cavity, or (b) that the shells are isolated SNRs apparently juxtaposed along 
the line of sight, but in fact spatially separated. \citet{N+01}, using a 
study of combined \rosat\ and \asca\ X-ray data, noted that the shells were 
spectrally distinct to an extent which would rule out the single-explosion 
scenario, and also concluded that, on the basis of its enhanced iron 
abundance, Shell A was probably the result of a Type Ia SN. 

In this paper, we present new \chandra\ observations of DEM\,L\,316.  
These observations allow us, for the first time, to spatially resolve 
the structure of the DEM\,L\,316 remnants, with reasonable spectral
resolution over smaller regions. We supplement these data with \xmm\
observations, which allow us to more reliably determine X-ray spectra 
for larger regions.  \S2 describes our observations, while \S3 contains 
our analysis.  A discussion of the implications of our findings is given 
in \S4, and those findings are summarized in \S5.

\section{Observations}

We observed DEM\,L\,316 with the \chandra\  Advanced CCD Imaging 
Spectrometer (ACIS) S3 back-illuminated chip (sequence number 500279, 
observation 2829, 40 ks).  Data were reprocessed following procedures 
recommended by the Chandra X-ray Center (CXC): we removed the afterglow 
correction, generated a new bad pixel file, and applied corrections for 
charge-transfer inefficiency (CTI) and time-dependent gain for an 
instrument temperature of -120 C. We also applied background cleaning 
using the 5$\times$5 pixel islands of our VFAINT mode observation. The
dataset was filtered for high-background times and poor event grades,
resulting in a total ``good time" interval of 35.6 ks, and restricted 
to the energy range of 0.3-8.0 keV, where the S3 chip is most sensitive.

Spectral results were extracted from the new event file. Background 
regions were taken from source-free areas surrounding the SNRs, and the
spectra of these background regions were scaled and subtracted from the
source spectra.  Individual spectra for regions of interest (Table~\ref{tab:regions}), and the corresponding primary and auxiliary
response files, were extracted with the \textsc{ciao} acisspec script 
and analyzed in \textsc{xspec}. Spectra were rebinned by spectral energy 
to achieve a signal-to-noise ratio of 3 in each bin.

\begin{figure}
\epsscale{0.6}
\plotone{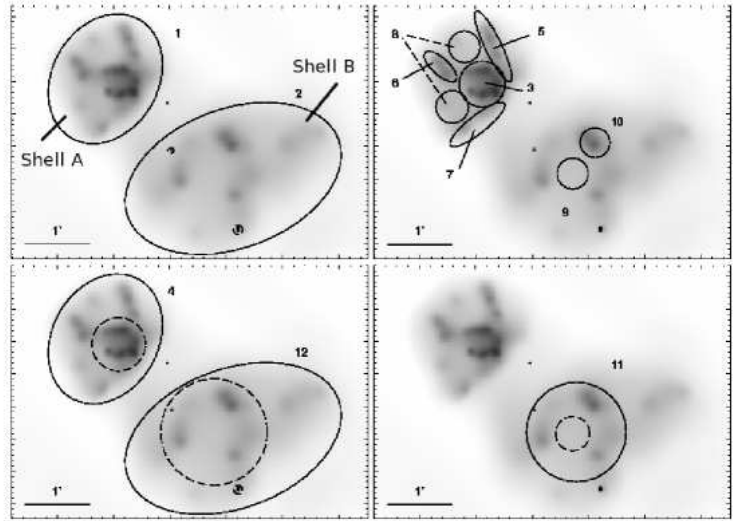}
\caption{Adaptavely smoothed \chandra\ ACIS image with spectral regions 
marked and labeled as listed in Table~\ref{tab:regions}. Regions with 
dotted lines are excluded from larger regions.  Excluded point sources
are also shown with dotted lines.}
\label{fig:reglog}
\end{figure}

Regions within DEM\,L\,316 were chosen to examine specific morphological 
features seen in our X-ray images (Figure~\ref{fig:reglog} and Table~\ref{tab:regions}). Regions 1 and 2 enclose the approximate X-ray 
boundaries of Shell A and Shell B, respectively.  Region 3 covers the 
bright central emission within Shell A, while Region 4 includes the 
limb of Shell A, excluding that central emission. Regions 5-7 cover 
brighter X-ray ``arcs" to the north, east, and south of Shell A; Region 
8 samples emission between those arcs.  In Shell B, Region 9 encloses 
the central emission from the remnant, and Region 10 covers a particularly 
bright knot of interior emission.  Region 11 incorporates the ring-like 
structure of brighter emission in Shell B, including the bright knot of
Region 10.  Region 12 covers emission on the limb of Shell B, excluding 
the center and the ring.

To augment our spectral data for Shells A and B (Regions 1 and 2)
we used \xmm\ EPIC MOS and pn data from observation ID 0201030101,
with exposure times of 11.2 ks (MOS1 and MOS2 detectors) and 
7.0 ks (pn detector). The pipeline-processed data obtained from
the Science Operations Centre (SOC) were reduced using the Science 
Analysis Software package provided by the SOC. The data were
filtered for poor event grades. Spectra for spatial regions 
corresponding to Regions 1 and 2 in the \chandra\ observations,
and surrounding background regions, were extracted from the filtered 
event files and rebinned to a minimum of 25 counts per bin.
Joint fits were performed in \textsc{xspec} to the data from 
\chandra's ACIS and the \xmm\ instruments.

For comparison, we utilized optical emission-line (\ha, \sii, \oiii ) 
images  taken with the Curtis-Schmidt telescope at Cerro Tololo Inter-American Observatory (CTIO), and radio images from the 
Australia Telescope Compact Array (ATCA) at 4.4 and 6.0 GHz 
(7 and 5 cm).  The angular resolution of the CTIO images was
about 1\farcs5; the half-power beamwidth of the radio images
was 12\arcsec. Other details of these observations are given 
in \citet{W+97}.  

\section{Results}

\subsection{X-ray Morphologies}

Previous X-ray analyses have noted a separation between the X-ray 
emission from the shells of DEM\,L\,316. In the \chandra\ ACIS data,
we indeed see two distinct shells; any overlap is too close to 
background levels to constitute a significant detection. Although 
diffuse emission between the shells is seen in the \xmm\ observations, 
this emission is faint ($\sim2\sigma$ above the background with the 
EPIC pn); in addition, the SNRs are over 7\arcmin\ off-axis, and so 
the spatial resolution may be insufficient to distinguish betwen the
boundaries of the two SNRs.

\begin{figure}
\epsscale{0.9}
\plotone{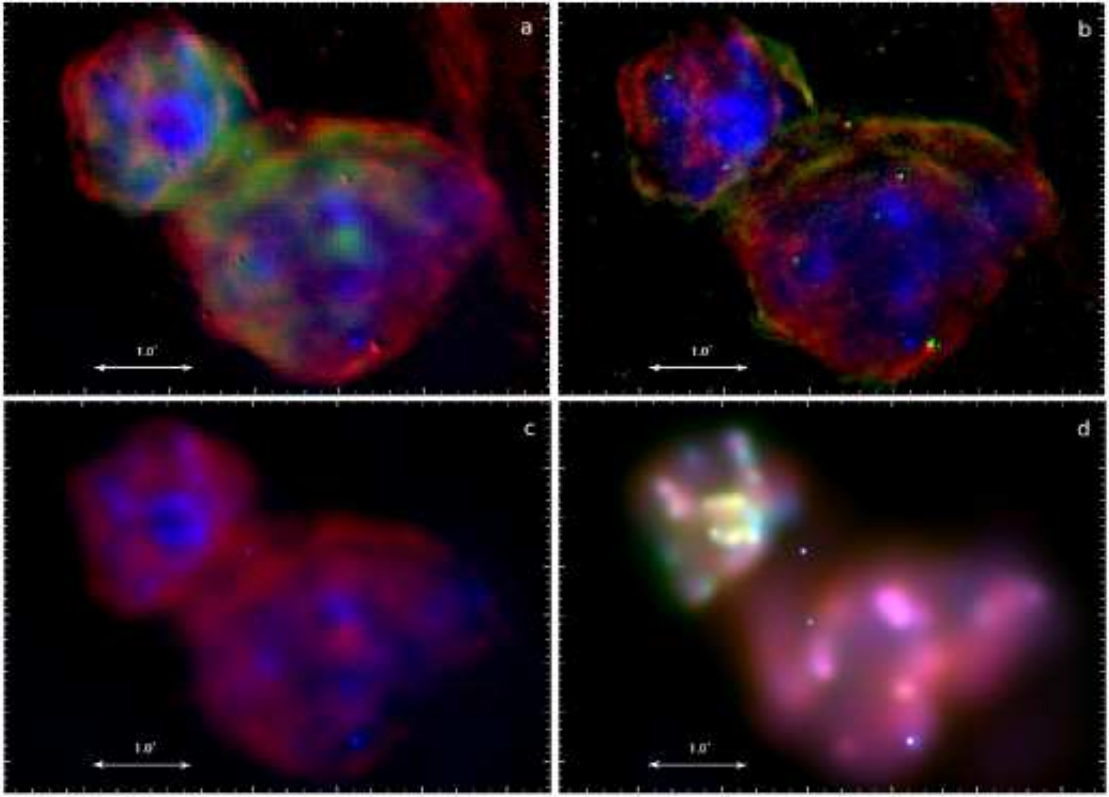}
\caption{(a) 3-band image showing \ha\ from the CTIO 
Curtis-Schmidt telescope (red), 7 cm radio map from the ATCA 
(green), and adaptively smoothed \chandra\ ACIS image (blue).  
(b) 3-band image showing \ha\ (red), \oiii\ (green), and 
smoothed ACIS image (blue). (c) 2-band image showing ATCA 7 cm
(red) and smoothed ACIS image (blue).  (d) 3-color \chandra\ ACIS 
image showing emission at 0.3-0.8 keV (red), 0.8-1.5 keV (green), 
and 1.5-8.0 keV (blue).  The X-ray images were adaptively smoothed 
on the same scale.}
\label{fig:3color}
\end{figure}

Three-color images comparing the \chandra\ morphology to emission 
in \ha, \oiii, and 7 cm radio \citep{W+97} are shown in Figure~\ref{fig:3color}a-c.  In general, the brightest regions of 
X-ray emission appear to correspond to areas that are fainter at 
radio wavelengths. Conversely, the X-rays grow fainter toward the 
radio-bright limbs.  Similarly, the X-ray emission appears 
well-confined by the optical filaments along the limbs of both 
shells.  In Shell A, some filaments across the face of the SNR 
appear to correspond roughly to regions of fainter X-ray emission, 
but there is no strong anticorrelation.

Shell A appears to have a bright ring of inner emission, as well as 
bright ``arcs" extending radially outward from that ring.  As noted 
by \citet{W+97}, the peak of the X-ray emission, now seen as the 
edge of this bright ring, corresponds to the location where optical 
filaments (particularly well seen in \oiii) from Shell B overlap with 
Shell A. The southwestern side of the SNR is flattened somewhat toward 
Shell B, and appears slightly brighter on the flattened side. Notably, 
this flattening occurs just north of the apparent overlap between the 
two shells, running parallel to the optical filaments and radio-bright 
limb there.  

The faint, outermost emission from Shell B is irregular, and is 
elongated in the E-W direction.  The primary source of this elongation 
seems to be an extension on the western side, which is matched by similar 
extensions in the optical and radio.  The X-ray emission of Shell B is 
distributed over the face of the remnant, out to the boundary defined
by the thick annulus of the radio structure.  Corresponding to the 
inner radius of this radio annulus, there appears to be a brighter ring 
of X-ray emission, though this is not as prominent as that in Shell A.  
A bright X-ray knot is visible on the north side of this ring.  Curiously,
this knot is directly north of a ``small-diameter source" noted in radio
images by \citet{W+97}.

To examine the morphology as a function of energy, we created a 
three-color image (Figure~\ref{fig:3color}d) with emission divided 
into energy bands of 0.3-0.8 keV (soft), 0.8-1.5 keV (medium), and 
1.5-8.0 keV (hard).  The  middle band was selected to show the 
contribution from the blend of Fe L lines that dominate the X-ray 
emission in this energy range, though it should be noted that other 
lines such as the helium-like blends of Ne and Mg also contribute 
to this band. The overall structure is similar in all three bands.  
Emission from Shell A shows considerably more medium-band emission 
than does Shell B; in particular, the central ring of Shell A is 
bright at these energies.  In contrast, the X-ray Bright Knot in Shell B 
appears brightest in the hard band. A similar three-color map using
the \xmm\ data reproduces these primary features at lower angular 
resolution: the medium band is much brighter in Shell A than in Shell 
B; and the location of the X-ray knot in Shell B is brighter in the 
hard band than is any other portion of either remnant.

\subsection{Spectral Fits}

We fit non-equilibrium ionization (NEI) plane-parallel shock models (``vpshock" in \textsc{xspec})\footnote{Details and references for 
the vpshock model can be found at \\ 
\url{http://heasarc.gsfc.nasa.gov/docs/xanadu/xspec/manual/node40.html.}},
modified by photoelectric absorption, to determine the plasma 
parameters for each spatially-selected region.  The model parameters 
are the mean shock temperature $kT$,  ionization timescale 
($\tau$=$n_{\rm e} t_{\rm ion}$), elemental abundances \citep[given 
as fractions of the solar abundance values of][]{AG89}, and a 
normalization $A$ proportional to the distance and volume emission measure 
($A ({\rm cm}^{-5}) = 10^{-14} \int n_{\rm e} n_{\rm H} dV/4\pi D^2$). 
Here,  $n_{\rm e}$ is the electron density, $n_{\rm H}$ the hydrogen 
density, $t_{\rm ion}$ the shock age (time since the earliest material 
was shocked), $V$ the volume occupied by the hot gas, and $D$ the 
distance, all in cgs units.

\subsubsection{Overall characteristics of the two shells}

We extracted spectra from regions covering all of Shell A and all 
of Shell B  (Table~\ref{tab:regions}, Regions 1-2) from both the 
\chandra\ and \xmm\ data. Spectra from the various instruments
were fit jointly. The fitted spectral parameters are given 
in Table~\ref{tab:shellspec}. It was not possible to obtain a 
statistically acceptable fit (at or above the 90\% confidence level) 
to both shell spectra with the same model parameters. 

The fitted absorption column densities were 3.6$\pm0.6\times10^{21}$ 
H-atom cm$^{-2}$ for Shell A, and 2.2 $\pm0.5\times10^{21}$ cm$^{-2}$ 
for Shell B under a range of model combinations.  Fits to \chandra\ 
data alone, however, gave absorption columns of 3.4$\pm0.2\times10^{21}$ 
cm$^{-2}$ for both remnants. The value obtained from joint fits for 
Shell B is similar to that of 2.0--2.5 $\times10^{21}$ cm$^{-2}$
estimated from fits to \rosat\ data by \citet{W+97}, although the value 
for Shell A is somewhat higher. The Shell A value, however, is reasonably
consistent with the estimate of 3.8$\times10^{21}$ cm$^{-2}$ that those
authors determined using \hi\ data for the LMC \citep{R+84} and a Galactic
foreground of 5$\times10^{20}$ cm$^{-2}$ \citep{SI91}.  Data from a 
more recent \hi\ survey of the LMC gives a higher figure of
6.2$\times10^{21}$ cm$^{-2}$ \citep{SK+03,KS+03} toward DEM\,L\,316, 
which when added to the aforementioned Galactic foreground gives a 
total absorption column of 6.7$\times10^{21}$ cm$^{-2}$.  Note that 
the \hi\ estimates include contributions from gas behind DEM\,L\,316.  
Conversely, contributions to X-ray absorption from, e.g., molecular 
gas will not be accounted for in \hi.
 
Shell A is well fit by a single vpshock component with high iron 
abundance.  Acceptable fits could be achieved with iron abundances 
between 1 and 5 times solar values; in all cases, substantially above 
the mean LMC abundance for iron.  The Mg and Si He-like line blends 
are prominent in the spectrum, and fitted abundances for Mg, Si and 
S were well above typical values for the LMC ISM. The best-fit 
model yields an absorbed 0.3$-$8.0 keV flux of 6$\pm 2 \times 10^{-13}$ 
erg cm$^{-2}$ s$^{-1}$, an unabsorbed flux of 2$\pm 1 \times 10^{-12}$ 
erg cm$^{-2}$ s$^{-1}$, and a luminosity of 3$\pm 2 \times 10^{35}$ 
erg s$^{-1}$. The upper limit on an additional power-law component 
is an absorbed flux of 1$\times10^{-14}$ erg cm$^{-2}$ s$^{-1}$, or 
1.7\% of the total flux; inclusion of such a component, however, 
does not improve the fit.

For Shell B, a model with a single vpshock component provides 
only a poor fit.  A substantial improvement in the fit is possible 
by using two components - either a combination of vpshock and 
power-law models or two vpshock models with the same abundances.  
For example, an F-test between the single vpshock model and a 
combined vpshock and power-law model gave only a $10^{-13}$ 
probability of no improvement in the fit.  The combination 
gives acceptable fits for values of the photon index $\Gamma$ 
in the 1.6-1.8 range, within the typical range of photon indices
for pulsar-wind nebulae in SNRs \citep{G03}.  Fits to Shell B do
not require abundance fractions much above the LMC mean values.
Models including an iron abundance of 1.0 solar or above 
could be ruled out at the 90\% confidence level.   The 0.3-8 keV 
flux for this model is 8$\times10^{-13}$ erg cm$^{-2}$ s$^{-1}$ 
(2$\times10^{-12}$ erg cm$^{-2}$ s$^{-1}$ unabsorbed), implying a 
luminosity of 3$\times10^{35}$ erg s$^{-1}$.  The power-law 
component takes up $\sim$30\% of the absorbed flux.

The two-plasma fit gives an ionization-equilibrium component
with a temperature $\sim$0.6 keV, and a non-equilibrium component with
kT$>$3 keV.  The fitted abundances are close to typical LMC abundances.
The fluxes and luminosities for this model fit are the same, to one 
significant figure, as those for the combined plasma/power-law fit.
In this case the high-temperature component accounts for $\sim$50\% 
of the flux.

\subsubsection{Regions within the SNR}

We used the \chandra\ ACIS data to obtain spectra from smaller 
spatial regions within DEM\,L\,316 (Figure~\ref{fig:reglog}).   We 
fit the best-fit spectral model found for Shell A (\S3.2.1) 
to the various sub-regions within Shell A (Regions 
3-8).  For most regions (3-6,8) the emission did not significantly 
deviate from this model, and a joint fit (with only normalizations 
varied) produced statistically acceptable results.  The exception 
was the S Arc region (Region 7), along the ``flattened" side of the 
SNR.  Emission from this region had an excess of soft emission when 
compared to the best-fit model for Shell A, and the joint fit to 
Shell A and the S Arc was poor. If we assume the abundances to be the 
same for this region as for Shell A generally, the S Arc region requires 
a slightly higher temperature (kT$\sim$1.5) and lower ionization 
parameter for a good fit.

Similarly, we used the two-plasma fit to Shell B (\S3.2.1) to perform 
joint fits with Shell B and its sub-regions (Regions 9-12). All of these
regions were well described by the overall two-plasma-component Shell B 
fit, although the relative normalizations of the two components differ.  
The high temperature component accounted for $>70$\% of the unabsorbed 
flux for the Bright Knot (Region 10), $\sim$40\% of the flux for the 
Center Ring (Region 11) and $<30$\%  of the flux in the Limb and Faint 
Center (Regions 9 and 12).  

We attempted to obtain additional information from these spatially 
resolved regions by fitting their spectra individually. To reduce the 
number of free parameters, we fixed the absorption column density to
the mean value of 3.4$\times10^{21}$ cm$^{-2}$ found from fits to 
\chandra\ data for both shells, and fixed all plasma abundances 
except iron to 30\% of their solar values. Given the low number of 
counts in most regions, it is unsurprising that the fits are highly
uncertain, as seen from the broad errors on each fitted parameter.
 
For all of these regions, a simple power-law fit alone could be ruled 
out at or above the 90\% confidence level.  The best power-law fit, 
though still not statistically acceptable, was to the data from the 
Bright Knot region of Shell B (Table~\ref{tab:regions}, Region 10).  
As with Shell B overall, an acceptable fit to the Bright Knot region 
could be obtained with a combination of plasma and power-law, or two 
plasma models at different temperatures.

Broadly, the spectra of various regions within Shell A are quite similar. 
The temperatures are consistently within the range of 0.9-1.2 keV, and 
the fitted iron abundance within the range of 1.0-1.7 times solar values.  
The discrepancy between our fitted values for Fe abundance in the entire 
Shell A and sub-sections of this shell appears to be due to the 
difference in fitting procedure.  For Shell A itself, we allowed Mg, Si, 
S and Fe to vary, while for the sub-regions we only allowed Fe to vary.  
When we fix all abundances except Fe at 0.3 for Shell A,  we obtain Fe 
abundances of 1.7$\pm$0.2 solar. Allowing the S abundance, for example, 
to vary freely leads to a significant improvement of the fit (F-test 
probability for no improvement of 0.08) and a larger Fe abundance.  
The greatest differences between spectral fits to the various regions 
appear to be in the ionization timescale $\tau$, which is lower in 
regions along the remnant limb than toward the center. If we assume 
that most of the bright X-ray emitting gas is recently shocked, this 
would imply an increase in density toward the bright central region.

For Shell B, the temperatures of the best-fit model combinations are 
generally in the 0.6-0.8 keV range, with iron abundances less than 40\% 
of solar values.  The bright ring of emission (Region 11) is set apart 
from other areas of Shell B in requiring a high temperature or a second
spectral component to provide an acceptable fit. The knot of emission
embedded in this ring (Region 10) is also better described with the
addition of a second spectral component, but the number of counts in 
this region is insufficient to distinguish between one- and two-component
models at a high level of significance.

\section{Discussion}

\subsection{Physical Properties of the SNRs}

We analyzed the overall emission from these SNRs, to determine 
abundance ratios between key elements within the SNRs; 
compare our findings with those previously made using \asca\ 
and \rosat\ instruments; and refine our estimates of the physical 
properties of the SNRs based on the \chandra\ and \xmm\ results.

\subsubsection{Abundances and SNR Types}

Our findings confirm the striking difference, first analyzed by 
\citet{N+01}, between the spectra of the two shells of DEM\,L\,316.
Model fits to the spectrum for Shell A require an iron abundance
significantly higher than that typical for the LMC, while no such 
condition is required for (or indeed permitted by) the spectrum of 
Shell B. The iron abundance found for Shell A, of three times solar 
abundances, is somewhat higher than the value of $\sim$1.9 solar 
found by \citet{N+01}.  However, as discussed in \S3.2.2, we can 
attribute this difference to the differing methods of spectral 
analysis.  The fits of \citet{N+01} fixed all elemental abundances 
except Ne and/or Fe to the mean LMC values of 0.3 solar.  When 
we use the same procedure in fitting the spectrum of Shell A, 
the fitted Fe abundances of 1.7$\pm$0.2 solar are within the 
error range of the \citet{N+01} value. 

The abundances found for Shell B are all well below the solar 
values, and in many cases near the typical LMC values 
\citep{RD92}.  Our values for the iron abundance in the entire
Shell B, and for sub-sections of this shell, all fall within 
the range of 0.16-0.4 solar.  \citet{N+01} did not fit the Fe 
value directly for this shell, but noted that a fit with a fixed 
Fe abundance of 0.3 solar provided a reasonable spectral fit.
These low abundance values suggest that the SNR is dominated by 
emission from swept-up material. However, the fitted values for
O and Ne are sufficiently above their LMC values, even considering
the large error ranges, to suggest that ejecta contributions to
the spectrum remain significant for some elements. This case is 
far from unique, as ejecta enrichment is still detectable in other
middle-aged LMC SNRs \citep[e.g.,][]{HBR03}.

The relative contribution of the lines from iron (compared 
with such elements as oxygen and neon) to the X-ray spectrum 
of a SNR is frequently used to discriminate between remnants
resulting from Type Ia versus Type II supernovae
\citep[e.g.,][]{HBR03,N+01,H+95}. Theoretical models 
\citep[e.g.,][]{B+03,I+99} of Type Ia SNe and their remnants 
predict substantially higher ratios of Fe to other elements than 
those seen in remnants of Type II SNe \citep[e.g.,][]{I+99,N+97}.  
A comparison of the ratios derived from Shells A and B to ratios
from SN models is shown in Table~\ref{tab:ratios}.

Using the abundances determined from the joint \chandra\ and \xmm\ 
fits (Table~\ref{tab:shellspec}) we find an overall O/Fe ratio of 
$\sim$1.5, and an Ne/Fe ratio of $\sim$0.2 for Shell A.  Models 
of Type Ia nucleosynthesis yields from \citet{I+99} give typical 
O/Fe ratios $\leq$1.0 and Ne/Fe ratios $\leq$0.1. In contrast, 
modeled yields from Type II SNe from \citet{N+97}, for progenitor 
masses of 15--40 $M_{\sun}$, give O/Fe in the 8--370 range and 
Ne/Fe in the 0.6--30 range.  Thus, the high iron abundance, 
particularly when contrasted with the less-enhanced Mg and Si, 
classes Shell A among LMC SNRs such as 0548-70.4 and 0534-69.9 
\citep{HBR03} and DEM\,L\,71 \citep{H+03}, which are tentatively 
categorized as remnants of Type Ia SNe due to high iron abundances.
(Supporting this categorization, SNR 0548-70.4 and DEM\,L\,71 also 
have Balmer-line dominated optical emission, a classic signature of 
a Type Ia remnant.)  We therefore agree with the conclusion of 
\citet{N+01} that Shell A is probably a remnant of a Type Ia SN.

For Shell B, depending on the model combination, the O/Fe ratios 
are in the range 30--130, and the Ne/Fe ratios in the range from 
8--16.  Typical ISM abundance ratios, based on the LMC values of \citet{RD92}, would give O/Fe of 13 and Ne/Fe of 2.4. The 
notably higher ratios found in Shell B suggest that these 
abundances not simply those one would expect from the swept-up 
interstellar gas, but are instead significantly enhanced by 
the contributions from ejecta.  The abundance ratios for Shell B 
are clearly within the ranges predicted for Type II SNe by the 
models of \citet{N+97}.  We therefore conclude that Shell B is  
very likely to be the remnant of a Type II SN.

\subsubsection{Densities, Energies and Pressures}

The fitted parameters for the NEI plane-shock spectral models can be 
used to derive a number of physical properties for the SNRs.  For this 
purpose we assume that hydrogen and helium are fully ionized, and that 
$n_{\rm He}/n_{\rm H}=0.1$; thus, the total particle density is about 
1.92$n_{\rm e}$. We assume each SNR to be an ellipsoid with the 
line-of-sight dimension equal to the length of the minor axis, 
with volume $V$ and hot gas volume filling factor $f_{\rm hot}$. 

The spectrum of Shell A is reasonably consistent throughout the 
remnant, and can be well described by a single spectral component. 
We therefore use the overall spectral fit to Shell A as the basis for
our estimation of its physical properties.  For Shell B, the picture 
is complicated by the likely presence of more than one spectral 
component, and by variations over the face of the remnant.  To 
characterize the properties of its overall expansion, therefore, we
use the single-component spectral fit to the limb region (Region 12)
as the basis for our calculations.  From the normalization factor for
a fitted spectrum, we can estimate the electron density and mass of 
the hot gas by:

\[  n_{\rm e} = 3.89\times 10^7  D A^{\half} V^{-\half} f_{\rm hot}^{-\half} 
\ \ \rm{cm}^{-3} \]

\[ M_{\rm gas} =  1.17 n_{\rm e}  m_{\rm H}  V f_{\rm hot} \ \ \rm{g} \]

The density and the plasma temperature (here given in keV) can be used 
to derive the thermal energy and pressure of the hot gas, according to:

\[ E_{\rm th} = 4.60\times 10^{-9} n_{\rm e}  T_{\rm keV} V f_{\rm hot}
\ \ \rm{erg} \]

\[ P_{\rm th} =  3.05\times 10^{-9} n_{\rm e}  T_{\rm keV}
\ \ \rm{dyn\ cm}^{-2} \]

Our derived values are summarized in Table~\ref{tab:snrprop}, in
terms of $f_{\rm hot}$ where appropriate. Quoted errors are simply propagation of errors in fitted parameters and volumes, and do not 
take into account uncertainties due to the choice of spectral models 
and filling factor.
  
The properties dependent on $n_{\rm e}$ are affected by our choice 
of volume filling factor. For spherical Sedov expansion, we expect  
$f_{\rm hot} \approx$ 0.25 \citep{S59}.  Using this value, we find 
electron densities of 0.3$\pm$0.1 cm$^{-3}$ for Shell A, and 
0.28$\pm$0.03 cm$^{-3}$for Shell B.  These densities are consistent 
with the broad ranges found by \citet{W+97} from \rosat\ data.  The 
electron density found for Shell A is also consistent, within the 
errors, with the ion density (presumed equal to the electron density) 
found by \citet{N+01} from \asca\ spectra, although it should be noted 
that their derived density was based on a value of $f_{\rm hot}=1$ for 
this shell.  If we similarly presumed $f_{\rm hot}=1$ to reflect the 
filled morphology, our density is only 0.2$\pm$0.1 cm$^{-3}$ for 
Shell A. For Shell B, the value we find from the SNR limb is somewhat 
lower than that found by \citet{N+01}; this difference is unsurprising, 
given that their spectral fit covered the entire remnant.

The thermal energies for the hot gas, using $f_{\rm hot} \approx$ 0.25,
are 1.5 $\pm$ 0.7 $\times 10^{50}$ erg and 2.1 $\pm$ 0.2 $\times 10^{50}$ 
erg for Shells A and B, respectively.  These values are also within the 
broad ranges found by \citet{W+97}. The pressures found with this filling
factor are 1.4 $\pm$ 0.6 $\times 10^{-9}$ dyn cm$^{-2}$ for Shell A and 
6.8 $\pm$ 0.7 $\times 10^{-10}$ dyn cm$^{-2}$ for Shell B, somewhat
higher than the \rosat-derived values.  We attribute this to the higher
temperature (or high-temperature component) found from \chandra\ and 
\xmm\ data; it is unsurprising that \rosat, sensitive in the 0.2-2.4 keV
range, would not have provided sufficient information at the high-energy 
end of the spectrum for accurate temperature determinations. (However, 
note that a higher filling factor for Shell A can lower the presumed
pressure.) Our findings reenforce the conclusion of \citet{W+97} that 
the hot gas thermal pressure exceeds the sum of the magnetic pressure 
and the thermal pressure in the warm ionized gas seen in optical emission
lines. 

\subsection{Spatially Resolved Features of the Two SNRs}

\subsubsection{Shell A}

Shell A fulfills many of the criteria for a ``mixed morphology" 
(also known as ``thermal composite") SNR \citep{RP98}: centrally 
brightened thermal X-ray emission, shell-like radio emission, and 
the absence of a prominent central compact source.  The complex 
internal X-ray morphology does not preclude this interpretation, 
as structures similar to the inner ring of emission in Shell A have 
been seen in other mixed morphology SNRs, such as Kes 79 and HB 9
\citep{RP98}. 

The identification of Shell A as a Type Ia SN  might initially 
appear to contradict the mixed morphology interpretation, as these 
SNRs are usually associated with massive-star phenomena.  Mixed 
morphology SNRs are often found in the vicinity of molecular clouds, 
as seen in their association with maser activity \citep[e.g.,][]{Y+03}. 
Several mixed morphology SNRs also contain pulsars or PWN \citep[e.g.,][]{W+05,S+04,R+94}.  However, it is also not 
uncommon for SNRs from Type Ia SNe to be seen in proximity to 
massive-star phenomena.  For example, the LMC SNR N103B, which 
shows X-ray inferred abundances suggestive of a type Ia 
origin\footnote{This conclusion is disputed; see \citet{V+02}.}
\citep{L+03,H+95}, is found at the periphery of \hii\ region 
DEM\,L\,84, within 40 pc of the NGC 1850 superbubble \citep{CK88}.
Although not itself a mixed morphology SNR, N103B demonstrates
that the association of many mixed morphology SNRs with 
massive-star phenomena does not rule out the possibility that
such a SNR might have a Type Ia origin.

Spectra of resolved regions for Shell A are remarkably similar 
in temperatures and abundances.  This isothermal structure is 
similar to that seen in Galactic mixed morphology remnants such 
as Kes 79 \citep[e.g.,][]{SS+04} and W44 \citep{S+04,R+94}.  
The lack of large temperature gradients across SNRs are often  
attributed to thermal conduction and/or turbulent mixing
\citep[e.g.,][]{SS+04,S+04,S+99}. To see whether the former 
possibility is plausible, we examine the timescale for thermal 
conduction \citep{S88,S62}:

\[ t_{\rm tc} \approx \frac{k n_{\rm e} l^2}{K_s} 
= 16,000\ {\rm yr} \left( \frac{n_{\rm e}}{1\ {\rm cm}^{-3}} \right)
\left( \frac{ l}{10\ {\rm pc}} \right)^2  
\left( \frac{kT_{\rm e}}{1\ {\rm keV}} \right)^{-5/2} 
\left( \frac{\ln \Lambda}{33} \right) \]

where $K_s$ is the \citet{S62} term for thermal conductivity
($1.8 \times 10^{-5}\ T^{5/2}/\ln \Lambda $ erg cm$^{-1}$ s$^{-1}$ 
K$^{-1}$), and $l$ the characteristic length scale for temperature 
variations ($T_{\rm e}/\vert \triangledown T_{\rm e} \vert $).
The Coulomb logarithm $\ln \Lambda$ is given by

\[ \ln \Lambda  = 32.2 + \ln \left( n_{\rm e}^{-1/2} 
\frac{kT_{\rm e}}{1\ {\rm keV}} \right) \]

which is $\sim$33 for our overall Shell A values of 
$n_{\rm e}=0.3$ and $T_{\rm e} \approx T = 1.4$ keV.
We compare this estimate for the thermal conduction 
timescale ($\le$ 16,000 yr) to estimated ages for the SNR.
\citet{W+97} suggest an age for Shell A of 27,000 yr, based on 
expansion velocities measured from echelle spectroscopy, and 
assuming Sedov expansion. Using the ionization timescale 
from X-ray spectral fits, \citet{N+01} estimated a larger 
age of $\sim$39,000 yr.  If the presence of magnetic fields 
does not significantly hinder thermal conduction, then, it is 
reasonable to expect temperature equilibration over the SNR.

One of the primary factors that, it has been argued, may 
disrupt thermal conduction is the tangling of the SNR's magnetic
field by internal mixing. However, in that case, turbulent mixing 
(i.e. bulk motions in the gas) during the SNR's development could 
itself produce a fairly flat temperature profile \citep{C+99, S+99}.  
In addition, one would expect these motions to create greater 
uniformity in the distribution of elements and in the ionization 
state of the gas.  We would, therefore, still not expect dramatic 
spectral variations across the SNR if turbulent mixing is a prominent
influence.

\subsubsection{Shell B}

In many respects, Shell B has the characteristics of an older 
remnant. The sub-solar metal abundances indicate substantial 
contributions from swept-up material to the X-ray emission, 
although, as discussed in \S4.1.1, some ejecta contributions are 
still detectable. The low surface brightness of the X-ray emission 
and relatively slow expansion velocity are also consistent with 
this picture.  

As with Shell A, Shell B has some of the characteristics of 
mixed-morphology SNRs: it is generally shell-type in radio, lacks 
a bright outer limb in the X-ray regime, and has substantial internal
emission. However, the case for Shell B is not as clear-cut as for 
Shell A.  The surface brightness profile of Shell B is shallower 
than that of Shell A, and the interior emission is more patchy.  
The presence of multiple spectral components for Shell B, and the 
indication that these components may be due at least in part to 
spectral variations across the remnant, are in marked contrast to 
the isothermal structure discussed above.  

The spatially-resolved spectra indicate that the high-energy 
spectral component is most prominent within the Bright Knot of  
X-ray emission.  Given its proximity to the ``small diameter radio 
source" mentioned in \citet{W+97}, and the identification of Shell B 
as the result of a Type II SN, we must consider the possibility
that the hard emission may come from a pulsar-wind nebula (PWN) 
embedded within the thermally-emitting shocked material.  The 
high-temperature component of its X-ray spectrum is certainly 
consistent with the spectral properties of a PWN. However, the 
association of the X-ray knot with the radio source is uncertain, 
as the X-ray peak is offset from the radio peak by 13\arcsec, 
greater than can be accounted for by the radio half-power beam 
width of 12\arcsec. A search of the X-ray power spectrum of the 
\xmm\ EPIC pn data for that region shows no evidence for periodicity 
in the emission; however, the low count rate of this source would
make the detection of any pulsations very unlikely.

In the absence of timing information or other indicators, and with 
the possibility remaining that the high-energy emission could 
be thermal, we cannot rule out the possibility that the Bright Knot 
emission may be simply a part of the remnant structure.  For example, 
a dense clump of swept-up ISM could have been incorporated into the 
interior of the SNR. If this material were to evaporate into the hot 
cavity within the SNR, it could raise the surface brightness of 
high-temperature gas shocked during more energetic phases of the 
SNR's expansion \citep[e.g.,][]{WL91}, and thus appear as a knot
of higher-temperature X-ray emission.

\subsection{Implications for Collision Scenario}

Several scenarios have been put forward to explain the unusual 
bi-lobed morphology of DEM\,L\,316.  We can examine these scenarios
in light of our new findings.   As put forth by \citet{N+01}, we can
decisively rule out the hypothesis that this system is the result of a 
single explosion into a bi-lobed cavity, on the basis of the wholly 
different heavy element abundances found for the two shells.  
On the other hand, we certainly cannot rule out the possibility that
the two SNRs are physically separate, and simply superposed along 
the line of sight.

With regards to the central assertion of \citet{W+97} that DEM\,L\,316 
consists of a pair of colliding SNRs, the case for collision is   
weakened by the tentative conclusion that Shell A is the remnant of a 
Type Ia SN and Shell B is the result of a Type II SN.  The explosions 
of two massive stars within the short lifetime of a SNR is conceivable 
if the two were members of a loose association, but if one SNR is the 
endpoint of the evolution of a low-mass star, no such association can 
be established.  Since the observable lifetime of a SNR is relatively 
short \citep[e.g.,][]{SC92},  the probability of two independent SNe 
from non-associated progenitors within that lifetime is low.   Even 
if the SN Ia progenitor had a main-sequence mass in the B-star range
\citep{W90}, the required timescales for stellar evolution to 
culminate in a Type Ia versus a Type II SN are sufficiently different 
to make the probability of collision between their SNRs quite low.  
Thus, such a collision must be presumed to be a rare event, and the 
chances of observing it small. 

Interpretation of the X-ray morphology of the remnants remains 
ambiguous.  Models of shock collisions often suggest that the hot 
cavities of two remnants undergoing a collision will be eventually 
connected by a ``tunnel" allowing hot gas to flow from one to 
another, but this connection may not form until relatively late 
times compared to the SNR observable lifetimes \citep{B+84,J+79,I78}. 
Based on the clear spectral differences between the shells, we can 
eliminate the scenario whereby the hot cavities are interacting, 
although this leaves open the possibility of a collision observed 
before such a tunnel has developed. In most cases the time between 
a SNR collision and the formation of a tunnel will be relatively
short, so the liklihood of observing a system in this state would
be small. However, if the SNRs collide during the late (radiative) 
stage of their evolution, such a tunnel may never form \citep{I78}.
Such collisions would primarily be characterized by the dense ``wall"
of cool material along the interface between the SNRs.
On the other hand, more recent laboratory and numerical simulations 
of colliding shocks \citep{V+01} show that reflected shock waves from 
the collision of two shocks will push the cavities of hot shocked 
gas apart from one another, producing density structures reminiscent 
of the observed multiwavelength morphologies of DEM\,L\,316.  These 
reflected shocks would be expected initially to produce X-ray 
emission in the ring of dense gas along the interface; however, the 
very high density expected in this region also leads to rapid cooling 
of the gas, with a correspondingly rapid drop in the observable X-ray 
emission.  In short, the \chandra\ observations do not provide any 
new support for the colliding-SNR scenario, but the scenario cannot
be conclusively ruled out on the basis of these data.

\section{Summary}

\chandra\ ACIS images and spectra, supplemented by \xmm\ spectra, 
reveal new details of the DEM\,L\,316 SNRs.  

\begin{enumerate}

\item Shell A exhibits a complex X-ray morphology, with a bright 
inner ring and ``arcs" of brighter emission embedded in an elliptical
region of diffuse emission. The brightest part of this ring 
occurs near an overlapping optically-emitting filament from Shell B.
The non-limb-brightened X-ray structure is in marked contrast to the
shell-like radio and optical morphologies, suggesting similarities 
to ``mixed morphology" SNRs.  As with many mixed morphology SNRs, 
little temperature variation is seen across Shell A. We suggest that
the favored explanations for such homogeneity in mixed morphology
SNRs, thermal conduction and turbulent mixing, are also applicable
to Shell A.
\item Shell B also has its brightest emission in a ring-like structure
well interior to the remnant limb.  In particular, a small knot on 
this ring shows the brightest emission in the SNR.  These regions are
also spectrally distinct from the SNR limb, as they require the addition
of a second, high-energy spectral component. The association of this
high-energy component with the X-ray knot suggests the presence of an
embedded pulsar-wind nebula, and its spectrum is consistent with this
supposition, although we cannot rule out other possible explanations.
\item We confirm the findings of \citet{N+01} that the 
spectrum of Shell A requires a high iron abundance, and thus that 
this SNR likely resulted from a Type Ia SN.  We are able to extend 
their work by explicitly fitting abundances for both shells, and
comparing abundance ratios to those predicted by models of 
Type Ia and Type II SNe.  We find that the O/Fe and Ne/Fe
ratios for Shell A are consistent with a Type Ia origin, while 
those for Shell B are consistent with a Type II origin.  In the
latter case, the observed ratios are significantly above those 
expected from swept-up ISM alone.
\item The inferred physical properties of the hot gas are broadly 
typical of middle-aged SNRs.  The relatively high thermal pressures 
in the hot gas for both SNRs emphasize the continued importance of 
the hot interiors to the evolution of the remnants.
\item The observed spectral differences between the SNRs strengthen 
the argument of \citet{N+01} that the shells are not part of a single 
bipolar SNR.  We further observe that the spectrally-based SNR
classifications weakens the case for the SNRs to be colliding, 
although the evidence is still far from conclusive.

\end{enumerate}

\acknowledgements 
The authors thank the anonymous referee for comments which have helped
to improve this paper. RMW acknowledges support from SAO grant GO2-3096A.
This material is based on work supported by the National Aeronautics and
Space Administration under  NASA grant NNG05GC97G issued through the LTSA
program.


\begin{deluxetable}{llcccc}
\tablecaption{Spatial regions used for spectral fits}
\tablehead{
\colhead{} & 
\colhead{SNR} & 
\colhead{Region} & 
\colhead{Counts\tablenotemark{a}} &
Center (J2000 RA, Dec) &
radii ($''$)
}
\startdata
1 & Shell A & entire SNR & 6352 $\pm$ 91 & 05:47:21.4 $-$69:41:28 & 63 $\times$ 49 \\
2 & Shell B & entire SNR & 7085 $\pm$ 140 & 05:46:58.8 $-$69:43:00 & 104 $\times$ 64 \\
3 & Shell A & Bright Center & 2822 $\pm$ 56 & 05:47:19.0 $-$69:41:33 & 21 \\
4 & Shell A & Limb & 3024 $\pm$ 85 & 05:47:21.4 $-$69:41:28 & 63 $\times$ 49, $-$25\tablenotemark{b}\\
5 & Shell A & N Arc & 737 $\pm$ 31 & 05:47:16.9 $-$69:40:58 & 36 $\times$ 8 \\
6 & Shell A & E Arc & 460 $\pm$ 24 & 05:47:26.3 $-$69:41:16 & 19 $\times$ 8 \\
7 & Shell A & S Arc & 480 $\pm$ 26  & 05:47:19.8 $-$69:42:10 & 32 $\times$ 8 \\
8 & Shell A & faint & 576 $\pm$ 31 & 05:47:22.0 $-$69:40:58 & 15 (N) \\
	&	&	&	& 05:47:24.2 $-$69:41:53 & 15 (S) \\
9 & Shell B & Faint Center & 317 $\pm$ 22  & 05:47:02.8 $-$69:42:55 & 15 \\
10 & Shell B & Bright Knot & 711 $\pm$ 29  & 05:46:58.9 $-$69:42:27 & 14 \\
11 & Shell B & Center Ring & 3555 $\pm$ 79 & 05:47:02.2 $-$69:42:54 & 46, $-$16\tablenotemark{b}\\
12 & Shell B & Limb & 2714 $\pm$ 112 & 05:46:58.8 $-$69:43:00 & 104 $\times$ 64, $-$49\tablenotemark{b} \\
\enddata
\tablenotetext{a}{Background-subtracted counts in the 0.3-8 keV range, over 
 35585 s exposure time.}
\tablenotetext{b}{Region with some interior emission excluded; radius 
 of excluded  region given as -X.}
\label{tab:regions}
\end{deluxetable}

\clearpage

\begin{deluxetable}{l|c|ccccc}
\tabletypesize{\scriptsize}
\tablecaption{Best-fit spectral models to SNR shells (joint \xmm\ - 
\chandra\ fits)}
\tablehead{
\colhead{Parameter} & 
\colhead{Shell A} & 
\multicolumn{5}{c}{Shell B} 
}
\startdata
component  & vpshock & vpshock & \multicolumn{2}{c}{vpshock + power-law} &
	\multicolumn{2}{c}{vpshock + vpshock} \\[6pt]
N$_H$ (cm$^{-2}$) & 3.6$^{+0.6}_{-0.5} \times10^{21}$ & 
	2.3$^{+0.4}_{-0.2} \times10^{21}$ &
	\multicolumn{2}{c}{2.2$^{+0.2}_{-0.2} \times10^{21}$} &
	\multicolumn{2}{c}{2.1$^{+0.2}_{-0.2} \times10^{21}$} \\[6pt]
kT (keV) & 1.4$^{+0.3}_{-0.2}$ & 0.65$^{+0.03}_{-0.02}$  & 
	0.59$^{+0.03}_{-0.03}$ & \nodata & 0.57$^{+0.4}_{-0.4}$ 
	& 5$^{+5}_{-2}$ \\[6pt]
 O/O$_\sun$ & 0.22$^{+0.8}_{-0.7}$ & 0.7$^{+0.2}_{-0.2}$   & 
 	0.75$^{+0.07}_{-0.07}$  & \nodata &
	\multicolumn{2}{c}{0.4$^{+0.1}_{-0.1}$} \\[6pt]
 Ne/Ne$_\sun$ & 0.2$^{+0.4}_{-0.2}$  & 0.6$^{+0.1}_{-0.2}$   & 
 	0.9$^{+0.4}_{-0.2}$  & \nodata &
	\multicolumn{2}{c}{0.7$^{+0.1}_{-0.2}$} \\[6pt]
 Mg/Mg$_\sun$ & 0.8$^{+0.3}_{-0.3}$ & 0.4$^{+0.1}_{-0.1}$   & 
 	0.6$^{+0.2}_{-0.2}$  & \nodata &
	\multicolumn{2}{c}{0.7$^{+0.3}_{-0.1}$} \\[6pt]
 Si/Si$_\sun$ & 0.9$^{+0.4}_{-0.3}$ & 0.35$^{+0.9}_{-0.8}$   & 
 	0.5$^{+0.2}_{-0.2}$  & \nodata &
	\multicolumn{2}{c}{0.5$^{+0.3}_{-0.1}$} \\[6pt]
 Fe/Fe$_\sun$ & 2.7$^{+0.8}_{-0.5}$ & 0.10$^{+0.03}_{-0.02}$ & 
 	0.15$^{+0.05}_{-0.03}$  & \nodata &
	\multicolumn{2}{c}{0.23$^{+0.09}_{-0.05}$} \\[6pt]
$\tau$ (cm$^{-3}$ s) & 1.7$^{+1.2}_{-0.5} \times10^{11}$ &  
	$<5\times10^{13}$ & $<5\times10^{13}$ & 
	\nodata & $<5\times10^{13}$ & 1.3$^{+0.7}_{-0.4} \times10^{11}$\\[6pt]
$\Gamma$ & \nodata & \nodata & \nodata & 1.7$^{+0.4}_{-0.2}$ & 
	\nodata & \nodata \\[6pt]
norm (cm$^{-5}$)&  2.1$^{+0.9}_{-0.7} \times10^{-4}$ & 
	1.7$^{+0.1}_{-0.1} \times10^{-3}$ &
	1.41$^{+0.03}_{-0.04} \times10^{-3}$ &
	5.3$^{+0.6}_{-0.5} \times10^{-5}$ &
	8.1$^{+0.4}_{-0.5} \times10^{-4}$ &
 	2.1$^{+0.2}_{-0.2} \times10^{-4}$  \\
$\chi^2_{\rm red}$ & 1.19 & 1.36 & \multicolumn{2}{c}{1.20} & 
	\multicolumn{2}{c}{1.17} \\
dof & 210 & 507 & \multicolumn{2}{c}{505} & \multicolumn{2}{c}{504} \\
\enddata
\tablecomments{Spectra cover the range between 0.3-8.0 keV. Elements 
not listed are fixed to a LMC mean abundance of 30\% solar \citep{RD92}.  
Quoted errors are the statistical errors in each fit parameter at the 
90\% uncertainty level.}
\label{tab:shellspec}
\end{deluxetable}

\begin{deluxetable}{lcccccc}
\tabletypesize{\scriptsize}
\tablecaption{Best-fit spectral models to regions in Shell A}
\tablehead{
\colhead{Parameter} & 
\colhead{Brt Ctr} & 
\colhead{Limb} &
\colhead{N Arc} & 
\colhead{S Arc} & 
\colhead{E Arc} & 
\colhead{Faint} 
}
\startdata
kT (keV) & 0.91$^{+0.01}_{-0.03}$ & 1.23$^{+0.07}_{-0.05}$ & 0.98$^{+0.09}_{-0.08}$ &
	1.1$^{+0.1}_{-0.1}$ &  1$^{+2}_{-0.2}$ & 0.91$^{+0.09}_{-0.01}$ \\[6pt]
 Fe/Fe$_\sun$ & 1.7$^{+0.4}_{-0.3}$  & 1.4$^{+0.4}_{-0.2}$  &  1.2$^{+0.5}_{-0.4}$ &
 	1.1$^{+0.8}_{-0.3}$ &  1$^{+2}_{-0.5}$ & 1.4$^{+0.9}_{-0.4}$ \\[6pt]
$\tau$ (cm$^{-3}$ s) & 4$^{+1}_{-3.6} \times10^{13}$ & 
	2.2$^{+0.4}_{-0.2} \times10^{11}$  & 1$^{+50}_{-0.4} \times10^{12}$ &  
	2$^{+1}_{-1} \times10^{11}$ & 6$^{+494}_{-5} \times10^{11}$ & 
	5$^{+4}_{-2} \times10^{11}$ \\[6pt]
norm (cm$^{-5}$)&  2.09$^{+0.07}_{-0.07} \times10^{-4}$ & 
	2.0$^{+0.1}_{-0.1} \times10^{-4}$& 6.4$^{+0.05}_{-0.04} \times10^{-5}$ &
	3.8$^{+0.03}_{-0.03} \times10^{-5}$ & 3.6$^{+0.3}_{-0.4} \times10^{-5}$ & 
	4.3$^{+0.3}_{-0.4} \times10^{-5}$ \\ 
$\chi^2_{\rm red}$ & 1.19  & 1.17 & 1.04 & 1.34 & 1.03 & 1.27 \\
dof & 91 & 177 &  54 & 44 & 36 & 57  \\
\enddata
\label{tab:regaspec}
\tablecomments{Here and in Table~\ref{tab:regbspec}, N$_H$ is fixed at 3.4$\times10^{21}$ cm$^{-2}$ and non-Fe abundances at 0.3 solar.  
The ``Limb" region incorporates the ``Arc" and ``Faint" regions.  All
model fits in this table use the vpshock model.}
\end{deluxetable}

\begin{deluxetable}{lcccccccccccccccc}
\tabletypesize{\scriptsize}
\tablecaption{Best-fit spectral models to regions in Shell B}
\tablehead{
\colhead{Parameter} & 
\colhead{Limb} & 
\multicolumn{2}{c}{Bright Ring} & 
\colhead{Faint Ctr} &
\multicolumn{2}{c}{Bright Knot}  
}
\startdata
component & vpshock & vpshock  & vpshock + pwrlw & vpshock &
	vpshock & vpshock + pwrlw \\[6pt]
kT (keV) & 0.78$^{+0.04}_{-0.05}$ & 1.8$^{+0.1}_{-0.2}$ & 0.63$^{+0.04}_{-0.03}$ &
	0.6$^{+0.1}_{-0.1}$ & 0.43$^{+0.09}_{-0.06}$  & 0.65$^{+0.08}_{-0.09}$ \\[6pt]
 Fe/Fe$_\sun$  & 0.29$^{+0.06}_{-0.07}$  & 0.31$^{+0.04}_{-0.05}$ & 
 	 0.16$^{+0.03}_{-0.04}$ & 0.1$^{+0.1}_{-0.1}$ & 0.4$^{+0.1}_{-0.1}$ & 
	 0.3\tablenotemark{a} \\[6pt]
$\tau$ (cm$^{-3}$ s)  & 1.8$^{+0.3}_{-0.3} \times10^{11}$  & 
	8$^{+1}_{-1} \times10^{10}$ &  6$^{+3}_{-1} \times10^{11}$  & 
	$<4 \times10^{13}$ &  $<4  \times10^{13}$ &  
	$<5  \times10^{13}$ \\[6pt]
$\Gamma$  & \nodata & \nodata & 2.2$^{+0.4}_{-0.2}$ & \nodata  & \nodata &
	2.6$^{+0.4}_{-0.4}$   \\[6pt]
norm (cm$^{-5}$) & 4.6$^{+0.3}_{-0.2} \times10^{-4}$ & 
	3.8$^{+0.1}_{-0.1} \times10^{-4}$ & 8.3$^{+0.4}_{-0.3} \times10^{-4}$  &
	1.1$^{+0.1}_{-0.2} \times10^{-4}$ & 5.1$^{+0.3}_{-0.4} \times10^{-5}$ &
	6$^{+1}_{-1} \times10^{-5}$  \\[6pt] 
norm2 (cm$^{-5}$) & \nodata & \nodata &  6$^{+1}_{-1} \times10^{-5}$ & 
	\nodata  & \nodata &  2.9$^{+0.4}_{-0.3} \times10^{-5}$   \\[6pt]
$\chi^2_{\rm red}$  & 1.22 & 1.26 & 1.05 &  0.80 & 1.29  &  1.08   \\
dof  & 268 & 189 & 187 & 35 & 61  & 60  \\
\enddata
\tablenotetext{a}{Could not be adequately constrained, so fixed to LMC 
mean abundance.}
\label{tab:regbspec}
\end{deluxetable}

\begin{deluxetable}{lcccccccccccccc}
\tabletypesize{\scriptsize}
\tablecaption{Element ratios for SNR shells}
\tablehead{
\colhead{Ratio} & 
\colhead{Shell A} &
\colhead{Shell B} &
\colhead{LMC\tablenotemark{a}} &
\multicolumn{2}{c}{\underline{\ Type Ia models\tablenotemark{b}} \  \ } &
\multicolumn{4}{c}{\underline{\hspace{1.4cm}Type II models\tablenotemark{c}\hspace{1.1cm}}} \\
\colhead{} &
\colhead{} &
\colhead{} &
\colhead{} &
\colhead{W7} &
\colhead{WDD1} &
\colhead{15M$_{\sun}$} &
\colhead{20M$_{\sun}$} &
\colhead{25M$_{\sun}$} &
\colhead{40M$_{\sun}$} 
}
\startdata
O/Fe & 1.5 & 30--130 & 13.2 & 0.67 & 0.46 & 8.48 & 66.0 & 178 & 367 \\
Ne/Fe & 0.2 & 8--16 & 2.4 & 0.016 & 0.0060 & 0.59 & 9.08 & 29.9 & 23.0 \\
O/Si & 5.9 & 19--48 & 3.5 & 1.63 & 0.57 & 8.18 & 25.8 & 45.1 & 30.6 \\
\enddata
\tablenotetext{a}{Typical LMC abundance ratios \citep{RD92}}
\tablenotetext{b}{Based on nucleosynthesis models for Type Ia SNe \citep{I+99}.
W7=fast deflagration model, WDD1=slow deflagration, delayed detonation model.}
\tablenotetext{c}{Based on nucleosynthesis models for Type II SNe \citep{N+97}.}
\label{tab:ratios}
\end{deluxetable}

\begin{deluxetable}{lcccccc}
\tabletypesize{\scriptsize}
\tablecaption{SNR properties from spectral fits}
\tablehead{
\colhead{} &
\colhead{Shell A} &
\colhead{Shell B} 
}
\startdata
$n_e$ (cm$^{-3}$) & 0.16 $\pm$ 0.07 $ f_{\rm hot}^{-\half}$ 
	& 0.14 $\pm$ 0.01 $ f_{\rm hot}^{-\half}$  \\
$M_{\rm gas}$ (M$_\sun$) & 50 $\pm$ 20 $ f_{\rm hot}^{\half}$ 
	& 110 $\pm$ 10 $ f_{\rm hot}^{\half}$  \\
$E_{\rm th}$ (erg) & 3.0 $\pm$ 0.9 $\times 10^{50}$ $ f_{\rm hot}^{\half}$ 
	& 4.2 $\pm$ 0.5 $\times 10^{50}$ $ f_{\rm hot}^{\half}$  \\
$P_{\rm th}$ (dyn cm$^{-2}$) 
	& 6.8 $\pm$ 0.9 $\times 10^{-10}$ $ f_{\rm hot}^{-\half}$ 
	& 3.4 $\pm$ 0.4 $\times 10^{-10}$ $ f_{\rm hot}^{-\half}$ \\
\enddata
\label{tab:snrprop}
\end{deluxetable}


\clearpage

\clearpage

\clearpage

\begin{figure}
\plotone{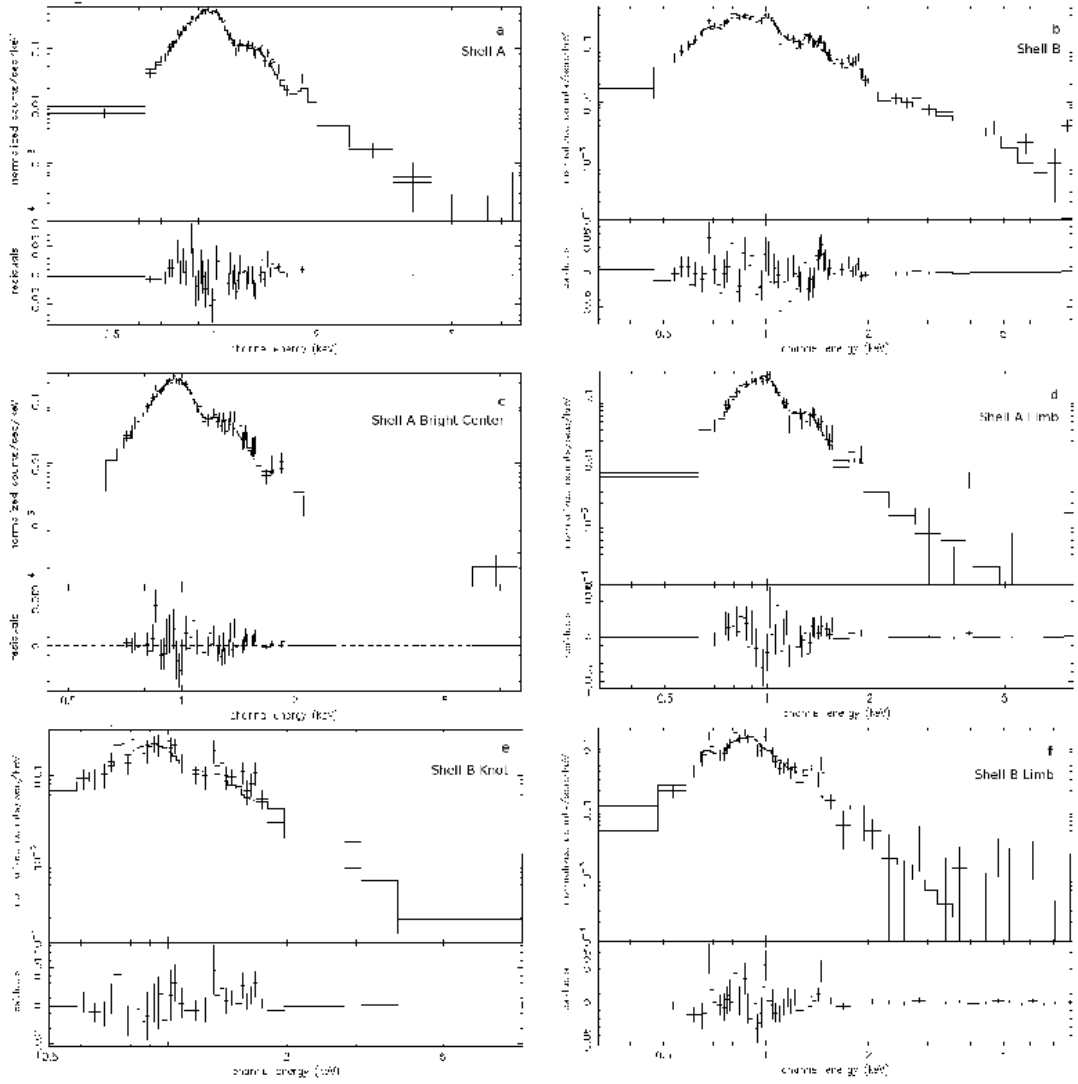}
\caption{Spectral fits to \chandra\ ACIS data for: (a) Region 1, 
Shell A; (b) Region 2, Shell B; (c) Region 3, Shell A Bright Center;
(d) Region 4, Shell A Limb; (e) Region 10, Shell B Bright Knot; and
(f) Region 12, Shell B Limb. }
\label{fig:spectra}
\end{figure}

\end{document}